\documentclass[letterpaper, 11pt, fleqn]{article}
\usepackage{amsfonts,epsfig,amsmath}
\usepackage{graphicx}

\begin{document}

\newtheorem{definition}{Definition}
\newtheorem{lemma}{Lemma}
\newtheorem{theorem}{Theorem}
\newtheorem{corollary}{Corollary}

\newcommand{\true}{\mbox{\sc true}}
\newcommand{\false}{\mbox{\sc false}}
\newcommand{\ox}{\overline{x}}
\newcommand{\oy}{\overline{y}}
\newcommand{\oz}{\overline{z}}
\newcommand{\amax}{\alpha_{\max}}
\newcommand{\gmax}{g_{\max}}
\newcommand{\keyw}[1]{{\bf #1}}
\newcommand{\e}{{\rm e}}
\newcommand{\whp}{w.h.p.}
\newcommand{\eps}{\epsilon}
\newcommand{\remove}[1]{}

\title{The Phase Transition in Exact Cover}

\author{Vamsi Kalapala and Cristopher Moore \\ {\tt \{vamsi, moore\}@cs.unm.edu}, \\ Department of Computer Science, \\ University of New Mexico, Albuquerque, NM 87131.}

\maketitle

 \begin{abstract}
We study EC3, a variant of Exact Cover which is equivalent to Positive 1-in-3 SAT.
Random instances of EC3 were recently used as benchmarks for simulations of an adiabatic quantum algorithm.  Empirical results suggest that EC3 has a phase transition from
satisfiability to unsatisfiability when the number of clauses per
variable $r$ exceeds some threshold $r^* \approx 0.62 \pm 0.01$.
Using the method of differential equations, we show that
if $r \le 0.546$ \whp\ a random instance of EC3 is
satisfiable.  Combined with previous results this limits
the location of the threshold, if it exists, to the range $0.546 < r^* < 0.644$.

\end{abstract}

\section{Introduction}

Numerous constraint satisfaction problems are believed to have a
``phase transition'' in the random case when the ratio $r$ of clauses
to variables crosses a critical threshold $r^*$: that is random formulas are \whp\ satisfiable if $r < r^*$, and \whp\ unsatisfiable if $r > r^*$, in the limit where the number of variables $n$ tends to infinity.  For 3-SAT, for instance, this ratio appears to be roughly $4.27$; see~\cite{tcs} for a review.

In this paper we study a similar phase transition in a variant of Exact Cover known as EC3~\cite{farhi0,wenjin}. An instance of Exact Cover consists of a set $S = \lbrace a_1, a_2, \ldots a_m \rbrace$ and a family of subsets of $S$, $F = \lbrace S_1, S_2, \ldots S_n \rbrace$. The problem is to determine whether there is a subfamily $C \subseteq F$ such that each element in $S$ is contained in exactly one $S_i \in C$.
In EC3, each $a_i \in S$ is restricted to appear in exactly three of the subsets $S_i \in F$.

EC3 can be formulated as Positive 1-in-3 SAT.  Here we have a set of boolean variables $V = \lbrace v_1, v_2, \ldots v_n \rbrace$ and a set of clauses $C = \lbrace c_1, c_2, \ldots c_m \rbrace$, where each $c_i \subset V \mbox { and } |c_i| = 3$. Note that the variables appear as positive literals only. A clause is satisfied when exactly one of its variables is true.
The problem is to determine whether any of the $2^n$ truth assignments satisfies every clause in $C$. An instance of EC3 can be transformed into an instance of Positive 1-in-3 SAT by setting
$v_i \leftarrow S_i$ and $c_i \leftarrow \lbrace v_j~|~a_i \in S_j \rbrace$. In what follows we refer to clauses and variables rather than sets and covers.

We conjecture that EC3 possesses a phase transition at some density $r^*$, where we construct random formulas with $m=rn$ clauses by choosing uniformly from among the ${n \choose 3}$ possible clauses with replacemment.  We note that techniques of Friedgut~\cite{friedgut} can be used to show that a {\em non-uniform} threshold exists, i.e., a function $r^*(n)$ exists such that, for any $\eps > 0$, random formulas are \whp\ satisfiable if $r < (1-\eps) r^*(n)$ and \whp\ unsatisfiable if $r > (1+\eps) r^*(n)$.
Interestingly, for 1-in-$k$ SAT where variables can be negated, Achlioptas, Chtcherba, Istrate and Moore~\cite{soda} established rigorously that a threshold exists, at $r^* = 1/{k \choose 2}$.

Knysh, Smelyanskiy and Morris~\cite{knysh} showed that random EC3 formulas are \whp\ unsatisfiable if $r \ge 0.644$, establishing an upper bound $r^* < 0.644$ if the transition exists.  Our main result establishes the lower bound $r^* > 0.546$.  Formally:
\begin{theorem}
Let $\phi$ be a EC3 formula consisting of $m = rn$
clauses chosen uniformly with replacement from the ${n \choose 3}$
possible clauses.  If $r < 0.546$, $lim_{n \to \infty} {\rm
Pr}[\phi \mbox{ is satisfiable}] = 1$.
\label{thm:lower}
\end{theorem}
Our proof uses the method of differential equations~\cite{worm} to show satisfiability with positive probability.
Satisfiability \whp\ then follows from the non-uniform threshold referred to above.

In addition to the transition phenomenon, our motivation is partly that Farhi et al. recently
simulated a quantum adiabatic algorithm~\cite{farhi1} on random instances of EC3.
They were only able to simulate this algorithm on
small numbers of variables (up to 17), but, in this range, the algorithm appeared to work
in polynomial time on formulas with a variety of values of $r$.  This
is exciting given that EC3 is NP-complete.  On the
other hand, van Dam and Vazirani~\cite{vv} showed that such algorithms
cannot succeed in polynomial time in the worst case, suggesting that
either the experiments in~\cite{farhi2} do not capture the asymptotic
behavior of the algorithm, or that random formulas are considerably
easier than worst-case ones.

\remove{
An interesting feature of EC3 is that we can
carry out exhaustive searches on surprisingly large random
formulas.  For 3-SAT, while state-of-the-art SAT
solvers such as Chaff can solve problems in VLSI verification of up to
1 million variables~\cite{chaff}, they can only
handle up to few hundred variables for random formulas due to their lack of structure.  For
EC3, on the other hand, we can solve random formulas
of up to about 1700 variables.  This allows us to obtain a rather sharp numerical estimate of the threshold $r^*$.
}

\section{The lower bound}

In this section we prove Theorem \ref{thm:lower} using the technique of differential equations.
Before delving into the proof, we first describe the mechanics
of setting variables in an EC3 formula. We call clauses of length $i$ in the formula
``$i$-clauses''.  $1$-clauses are also called unit clauses.
Setting a variable $v$ $\false$ replaces each 3-clause $t_i = \{ v, x_i, y_i \}$ it appears in with a 2-clause $x_i \oplus y_i$, and replaces each 2-clause $b_i =  v \oplus z_i$ it appears in with a positive unit clause $z_i$. Similarly, setting $v$ $\true$ replaces each 3-clause it appears
in with two negative unit clauses $\overline{x_i}, \overline{y_i}$, and replaces each
2-clause it appears in with a negative unit clause $\overline{z_i}$.

We analyze a simple greedy algorithm which is a variant of Unit Clause
resolution or UC for short~\cite{ChaoFranco}.
Algorithms based on UC have so-called ``free'' and ``forced'' steps. A free step is one in which
the algorithm decides on a variable and the value to which that variable is set.
Forced steps result from unit propagations, i.e., repeatedly satisfying all unit clauses until none are left.
Two of the common ways to choose the variable on the free step are
\begin{enumerate}
\item choose a variable at random,
\item for a fixed $i$, choose an $i$-clause at random, then choose a variable at random from
one of the $i$ variables in the clause.
\end{enumerate}
We obtained the best lower bounds by using method 2, known as Short Clause or SC, and always setting the chosen variable to $\true$.  Our algorithm is shown in table 1.

\begin{table}[t]
\label{algotab}
\noindent\mbox{}\rule{\textwidth}{1pt}
\begin{tabbing}

\keyw{while} \= there are any unset variables, \keyw{do} \{  \\
\>{\tt // Free step.} \\
\>\keyw{if}   th\=ere are any 2-clauses \\
\>\>choose a clause $c$ at random from the 2-clauses \\
\>\keyw{else}\= \\
\>\>choose a clause $c$ at random from the 3-clauses \\
\>choose a variable $x \in c$ at random\\
\>set $x = \true$ \\
\>{\tt // Forced steps.} \\
\>\keyw{while} there are unit clauses, satisfy them; \\
{\tt \} } \\
\noindent\mbox{}\rule{\textwidth}{1pt}
\end{tabbing}
\caption{Our Algorithm - SC.}
\end{table}

We call each iteration of the outer {\tt while} loop, i.e., a free step followed by a series of forced steps, a {\em round}.  Since resolving a unit clause creates more unit clauses, the
forced steps are described by a branching process.  Our main goal
will be to show that this branching process \whp\ remains subcritical
throughout the algorithm for sufficiently small $r$, so that
the number of variables set in any round will be $O(1)$ w.h.p.

To analyze our algorithm we need to track the change in the number
of 2-clauses and 3-clauses in each round. Note that at the start of the algorithm we have no 2-clauses, it can be shown that after $o(n)$ free steps, the number of 2-clauses is \whp\ positive, and returns to zero only after $\Theta(n)$ free steps. This can be proved similar to lemma 3 in~\cite{AchMolloy}, by showing that the expected number of 2-clauses is positive after $o(n)$ steps.  As will be shown later, once the 2-clauses are exhausted the remaining 3-clauses form a very sparse formula, which can easily be satisfied.  Therefore, we focus on the phase of the algorithm when \whp\ 2-clauses exist, in which case the free step always sets a variable in a 2-clause.

\remove{
of density less than $0.05$. At this density the graph of the clause to variable connectivity of the formula \whp\ does not contain a giant component (indeed, with positive probability this graph consists of trees only), a simple argument then shows that the remaining formula is satisfiable with positive probability.
}

In what follows we describe the branching process corresponding to the forced steps. We then analyze the expected effect of each round, and give a set of differential equations that describe the ``trajectory'' of the algorithm.  Finally, we solve these differential equations and show that for $r \le 0.546$ the branching process remains subcritical.

Let $n$ be the number of variables in the formula.
Let $m = rn$ be the number of clauses. Let $T = t \cdot n$ be the number of rounds
completed so far. For $i=2,3$ let $S_i(T) = s_i(t) \cdot n$ be
the number of clauses of length $i$. Let $X(T) = x(t) \cdot n$ be the number of variables
set so far. Let $m_T, m_F$ be the expected number of variables set to $\true$, $\false$
respectively in each round (inclusive of the variable set in the free step).

We compute $m_T, m_F$ according to a two-type branching process as in~\cite{4d3c}.
The two types here are positive and negative unit clauses.
In the free step we set a variable in a 2-clause to $\true$ and this forces us to set the other variable
in the 2-clause to $\false$. Thus the initial expected population of unit clauses can be represented by a vector
\begin{equation}
	p_0 = \left( \begin{array}{c} 1 \\ 1 \end{array}  \right)
\end{equation}
where the first and second components count the positive and negative unit clauses
respectively.

We wish to determine the transition matrix of the branching process. If $X$ variables
have been set so far, the probability of a variable appearing in a given $i$-clause
is $i/(n-X)$. So, setting a variable to $\true$, i.e., satisfying a positive unit clause, creates, in expectation, $(6S_3 + 2S_2)/(n-X)$ negative unit clauses. Similarly, satisfying
a negative unit clause creates, in expectation, $2S_2/(n-X)$ positive unit clauses. Thus,
we have the following transition matrix $M$ for the branching process:
\begin{equation}
	M = \frac{1}{n-X} \left( \begin{array}{cc} 0 &  6S_3 + 2S_2 \\
	                                           2S_2 & 0 \end{array} \right) =
				\frac{1}{1-x} \left( \begin{array}{cc} 0 & 6s_3 + 2s_2 \\
	                                           2s_2 & 0 \end{array} \right)
	                                           \enspace .
\end{equation}
As long as the largest eigenvalue $\lambda_1$ of $M$ is less than 1, the expected
number of variables set to true or false in each round is given by the geometric series
\begin{equation}
\left( \begin{array}{c} m_T \\ m_F \end{array} \right) = (I + M + M^2 + \ldots) \cdot p_0 = (I - M)^{-1}  \cdot p_0
\end{equation}
where $I$ is the identity matrix. Moreover, as long as $\lambda_1 < 1$ throughout the algorithm,
i.e., as long as the branching process is subcritical for all $x$, $m_T \mbox{ and } m_F$ remain $O(1)$ and, as in~\cite{delb,4d3c}, our algorithm succeeds with positive probability. On the other hand, if $\lambda_1$ ever exceeds 1, then the branching process becomes supercritical,  the unit clauses proliferate with high probability and the algorithm fails. Note that
\begin{equation}
	\lambda_1 = \frac{2}{1-x}\sqrt{s_2(s_2 + 3s_3)}
\end{equation}
Our next step is to write down the expected change in $S_2, S_3 \mbox{ and } X$ in a given round as a function of their values at the beginning of the round. We define $\Delta f(T) = f(T+1) - f(T)$. Then:
\begin{eqnarray}
E[\Delta X(T)] & = & m_T + m_F
\label{Xeqn}
\\
E[\Delta S_3(T)] & = & -(m_T + m_F) \frac{3S_3}{n-X} + o(1)
\label{S3eq}
\\
E[\Delta S_2(T)] & = &  m_F \frac{3S_3}{n-X} - (m_T + m_F) \frac{2(S_2 - 1)}{n-X} - 1 + o(1)
\label{S2eq}
\end{eqnarray}
To see this, recall that in expectation we set $m_T + m_F$ variables inclusive of the variable chosen on
the free step, giving~\eqref{Xeqn}.  Any variable set during a round appears in $3S_3/(n-X)$ 3-clauses and $2S_2/(n-X)$ 2-clauses, in expectation; these clauses are removed, giving~\eqref{S3eq} and first negative term in~\eqref{S2eq}, the $o(1)$ terms absorb the probability that a given clause is ``hit'' twice during a round. Among the 3-clauses, those that had a variable set to false become 2-clauses, giving the positive term in~\eqref{S2eq}.  Finally, the $-1$ in~\eqref{S2eq} comes from the fact that SC chooses a random 2-clause and removes it on the free step.

Wormald's Theorem~\cite{worm} allows us to rescale~\eqref{Xeqn},~\eqref{S3eq}, and~\eqref{S2eq} to form a system of differential equations for $s_i(x)$.  The random variables $S_i(xn)$ will then be \whp\ within $o(n)$ of $s_i(x) \cdot n$ for all $x$, where $s_i(x)$ are the solutions to these equations.  By changing the variable of integration to $x$ and ignoring $o(1)$ terms, we transform these equations to the following simpler form:
\begin{eqnarray}
\frac{ds_3}{dx} & = & - \frac{3s_3}{1-x}
\label{s3xeq}
\\
\frac{ds_2}{dx} & = & \frac{m_F}{(m_T+m_F)} \frac{3s_3}{1-x} - \frac{2s_2}{1-x} - \frac{1}{m_T+m_F}
\label{s2xeq}
\end{eqnarray}
The initial conditions are $s_3(0) = r, s_2(0) = 0$, even though as in~\cite{delb} the differential equations trace the evolution of $s_2$ and $s_3$ after a $o(1)$ fraction of the variables have been set.

When $r = 0.546$, numerically
solving the differential equations gives us, at $x \approx 0.29$, $\max_x(\lambda_1) \approx 0.996 < 1$, so the branching process remains subcritical. At $x \approx 0.79$, the density of the 2-clauses becomes $s_2(x) = 0$. This means the algorithm succeeds with positive probability in exhausting all the 2-clauses. The density of the remaining 3-clauses is $s_3(x)/(1-x) \approx 0.02$. For EC3 formulas with such low densities, the graph of clause to variable connectivity (i.e., the graph in which clauses are nodes and clauses that have a variable in common have an edge between them) with positive probability consists of trees only (and in the terminology of~\cite{knysh} the formula has no ``core''). The formula can then be satisfied by repeatedly satisfying variables on the leaves of these trees. As a result, the algorithm succeeds with positive probability whenever $r \leq 0.546$, completing the proof of Theorem~\ref{thm:lower}.

We analyzed two other kinds of free steps, but they gave weaker bounds.  Setting a random variable  $\true$ gives $r^* > 0.5097$, and choosing a random 3-clause and setting one of its variables $\true$ gives $r^* > 0.5386$.  Probabilistic mixes of these steps with SC also appear to give weaker bounds.

\section{Numerical experiments}

\begin{figure*}[h] 
\begin{center}
\begin{tabular*}{14.5cm}{ll}
\includegraphics[width=6cm, height=4cm]{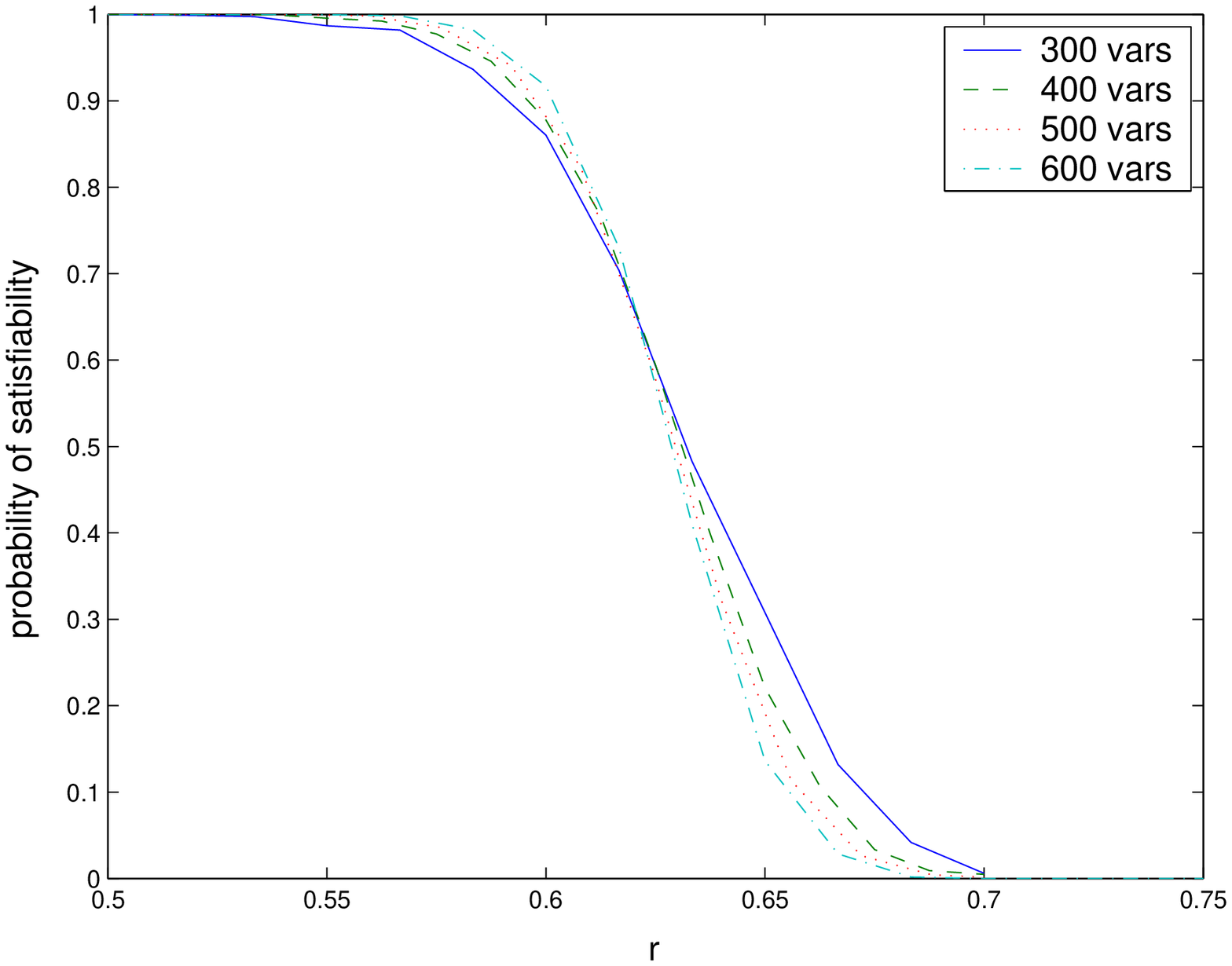} &
\includegraphics[width=6cm, height=4cm]{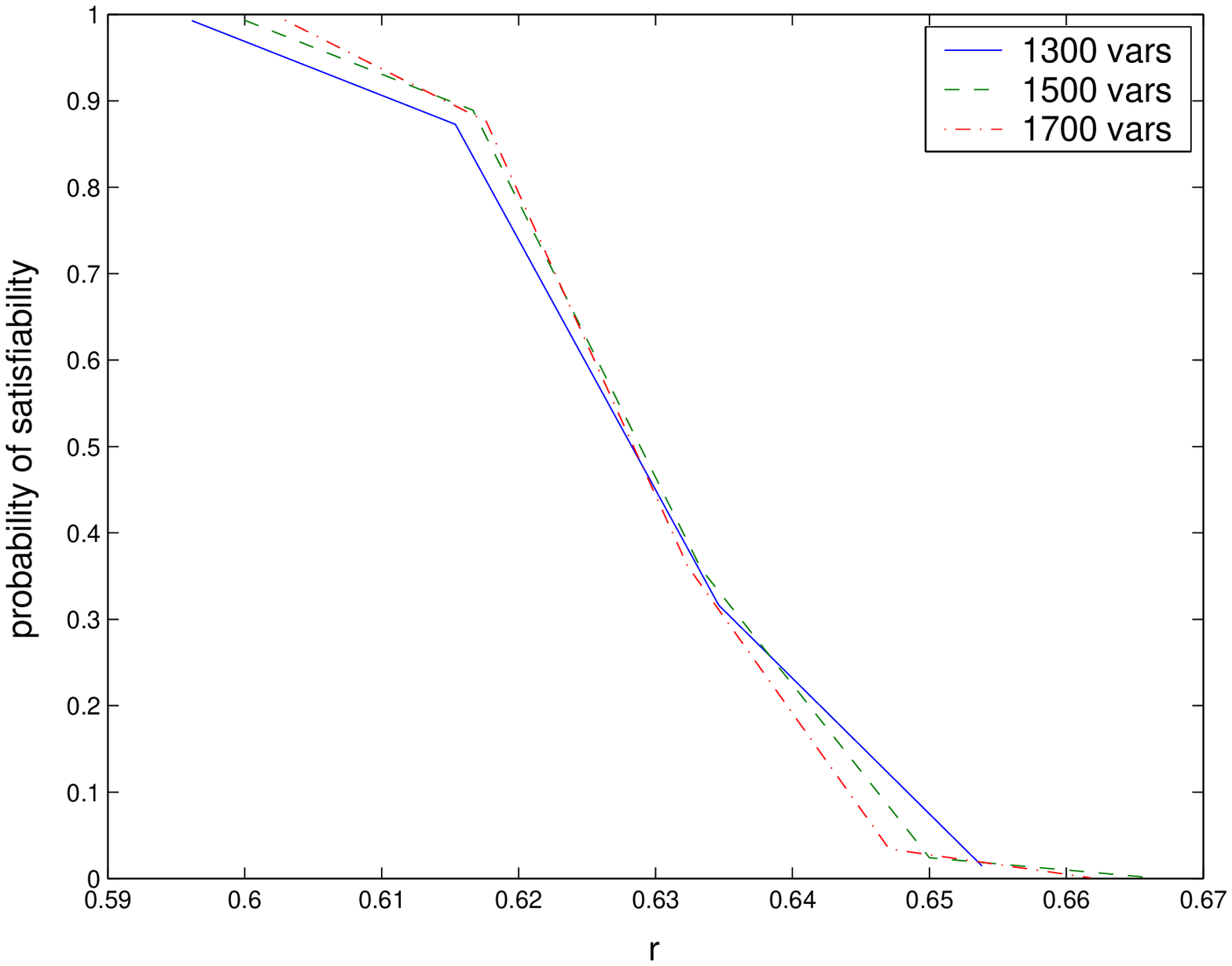} \\
\end{tabular*}
\end{center}
\caption{The probability of satisfiability as a function of $r$ for $n=300,400,500$ and $600$.}
\label{fig:pht}
\end{figure*}

We conclude with our own numerical experiments.
\remove{
In this section we describe the numerical experiments conducted to estimate the threshold of the phase transition in EC3. The clause to variable ratio $r^*$ about which the probability of satisfiability of a random EC3 formula drops from being close to 1 to being close to 0 is called the crossover point or threshold of the phase transition. In what follows, a random EC3 formula has $n$ variables and $m = rn$ clauses.

To estimate the threshold, we measure the probability of satisfiability of a random EC3 formula as a function of $r$ for various system sizes i.e., various values of $n$, and find a value of $r$ at which these curves appear to intersect. This is a standard approach in finite-size scaling for numerical experiments.
 }
For each value of $r$ and $n$ we performed $10^4$ trials, each of which consisted of creating a random EC3 formula and checking whether it is satisfiable or not using the 3-SAT solver Satz~\cite{ChuMinLi}.  The fraction of these which are satisfiable, as a function of $r$ for various values of $n$, is shown in Figure~\ref{fig:pht}.  Using the place where these curves cross as our estimate of the threshold (a common technique in finite-size scaling) suggests that $r^* \approx 0.62$.

\section{Conclusion} We have placed a lower bound of $r^* > 0.546$ on the threshold of the
phase-transition in EC3. Combined with the upper bound of $r^* < 0.644$~\cite{knysh}, a fairly small gap of $0.098$ remains.  It might be possible to improve our lower bound using algorithms that choose a variable based on the number of its occurrences in the remaining formula, as in~\cite{4d3c,Kaporis}.

\section{Acknowledgements}

The authors thank Sahar Abubucker and Arthur Chtcherba for reading the manuscript and providing valuable comments. This work was supported in part by NSF grants PHY-0200909, PHY-0071139 and the Sandia University Research Project.

\end{document}